\documentclass[11pt]{article}
\usepackage{amsmath,epsfig,wrapfig}
\begin{document}
\pagestyle{empty}

\begin{flushleft}
\Large
{SAGA-HE-156-99
\hfill December 27, 1999}  \\
\end{flushleft}

\vspace{2.0cm}
 
\begin{center}
\LARGE{{\bf Possible Studies of Parton Distributions}} \\
\vspace{0.2cm}

\LARGE{{\bf at JHF: polarized pd Drell-Yan process}} \\

\vspace{1.1cm}
 
\LARGE
{S. Kumano $^*$ }         \\
 
\vspace{0.4cm}
  
\LARGE
{Department of Physics}         \\
 
\vspace{0.1cm}
 
\LARGE
{Saga University}      \\
 
\vspace{0.1cm}

\LARGE
{Saga 840-8502, Japan} \\

\vspace{1.4cm}
 
\LARGE
{Talk given at the Workshop on} \\

\vspace{0.2cm}

{``Di-lepton Experiments at 50-GeV PS"} \\

\vspace{0.4cm}

{KEK, Japan, Oct. 29 -- 30, 1999 } \\

{(talk on Oct. 29, 1999)}  \\
 
\end{center}
 
\vspace{1.2cm}

\vfill
 
\noindent
{\rule{6.0cm}{0.1mm}} \\
 
\vspace{-0.3cm}
\normalsize
\noindent
{* Email: kumanos@cc.saga-u.ac.jp. Information on his research
       is available at} \\

\vspace{-0.4cm}
\noindent
{\ \ \  http://www-hs.phys.saga-u.ac.jp.}  \\

\vspace{+0.65cm}
\hfill
{\large to be published in proceedings}

\vfill\eject
\setcounter{page}{1}
%%%%%%%%%%%%%%%%%%%%%%%%%%%%%%%%%%%%%%%%%%%%%%%%%%%%%%%%%%%%%%%%%%%%%%%%%%%%%%
%%%%%%%%%%%%%%%%%%%%%%%%%%%%%%%%%%%%%%%%%%%%%%%%%%%%%%%%%%%%%%%%%%%%%%%%%%%%%%
\pagestyle{plain}
\begin{center}
 
\Large
{Possible Studies of Parton Distributions} \\

{at JHF: polarized pd Drell-Yan process} \\
 
\vspace{0.5cm}
 
{S. Kumano $^*$}             \\
 
{Department of Physics, Saga University}      \\

{Honjo-1, Saga 840-8502, Japan} \\

\vspace{0.7cm}

\normalsize
Abstract
\end{center}
\vspace{-0.60cm}

\begin{center}
\begin{minipage}[t]{10.0cm}
We discuss possible studies of parton distributions at the proposed
50-GeV PS facility.
First, comments are given in general on the importance of large-$x$
physics for finding new physics. Second, polarized deuteron reactions
are discussed in connection with new structure functions which are specific
for a spin-1 hadron. Third, if the polarized proton-deuteron Drell-Yan
process is studied at the facility, it could provide information
on the light-antiquark flavor asymmetry in the polarized distributions.
\end{minipage}
\end{center}

\vspace{0.1cm}
%%%%%%%%%%%%%%%%%%%%%%%%%%%%%%%%%%%%%%%%%%%%%%%%%%%%%%%%%%%%%%%%%%%%%%%%%%%%%%
%%%%%%%%%%%%%%%%%%%%%%%%%%%%%%%%%%%%%%%%%%%%%%%%%%%%%%%%%%%%%%%%%%%%%%%%%%%%%%
\section{Introduction}\label{intro}
\vspace{-0.1cm}

Although $Q^2$ dependence of parton distributions is
calculated by perturbative QCD and it has been confirmed by experiments,
the distributions themselves cannot be calculated without relying on
nonperturbative methods. Therefore, determination of the parton
distributions is important for testing nucleon structure models. 
However, it is also important for finding any exotic physics
signature in hadron reactions possibly beyond the current
theoretical framework.

It is a good idea to study such distributions at the proposed 50-GeV
facility in Japan. At this stage, secondary-beam experiments have been
mainly focused in the proposal, but there are not extensive studies
on primary-beam projects. However, it could be a valuable facility
in investigating large-$x$ parton distributions by using the primary beam.
In particular, if the proton polarization is attained, the facility
is a unique one for investigating polarized parton distributions
in the medium- and large-$x$ regions.  Although the RHIC-Spin project,
for example, investigates the spin structure of the proton, it measures
mainly on the smaller-$x$ region. In this sense, the 50-GeV facility
is compatible with the RHIC-Spin and other high-energy projects,
and it is important for understanding the nucleon structure
in the whole $x$ range.

Because the unpolarized antiquark flavor asymmetry,
nuclear structure functions, and the $g_1$ structure function
are discussed by other speakers \cite{ps},
the author would like to address himself to different topics.
In Sec. \ref{large-x}, comments are given on interesting large-$x$
physics. Then, polarized proton-deuteron (pd) Drell-Yan process is
discussed for studying unmeasured structure functions for
a spin-1 hadron in Sec. \ref{pd-dy}. Using a polarized pd
Drell-Yan formalism, we discuss the possibility of extracting
flavor asymmetry in polarized light-antiquark distributions
in Sec. \ref{flavor}. The summary is given in Sec. \ref{summary}.

\vspace{0.1cm}
%%%%%%%%%%%%%%%%%%%%%%%%%%%%%%%%%%%%%%%%%%%%%%%%%%%%%%%%%%%%%%%%%%%%%%%%%%%%%%
%%%%%%%%%%%%%%%%%%%%%%%%%%%%%%%%%%%%%%%%%%%%%%%%%%%%%%%%%%%%%%%%%%%%%%%%%%%%%%
\section{Comments on large-$x$ physics}
\label{large-x}
\vspace{-0.1cm}

There are three interesting topics on the large-$x$ physics as far as
the author is aware. First, the counting rule is usually used for
predicting parton distributions at large $x$, so that the large-$x$
measurements are valuable for testing the idea. Second, because nuclear
corrections are generally large in such a $x$ region, the experiments
provide important information on nuclear models at high energies.
This topic is related to the first one, e.g. in the studies of
$F_2^n/F_2^p$ \cite{mt}, because the deuteron and $^3$He targets are
used for measuring the neutron structure function. 
Third, it is crucial to know the details of the parton distributions
for finding new exotic signatures. 
The first two topics are discussed in other publications, so that
the interested reader may read for example Ref. \cite{mt}.
Because the third topic would be much important in relation to other
fields of particle physics, we discuss more details. 

In the recent years, anomalous events were reported at Fermilab
and DESY in the very large $Q^2$ region. We cannot judge precisely
whether or not these are really ``anomalous" in the sense that
the parton distributions, especially the gluon distribution, are not
well known in the large-$x$ region. For example. the CDF anomalous
jet data originally indicated that the perturbative QCD could not
explain the data. However, noting the gluon subprocesses play
an important role in the large-$E_T$ region, we could explain the data
by adjusting the gluon distribution at large $x$ \cite{cteq4}.
However, nobody knows that this is a right treatment because
there is no independent experiment for probing the gluon
distribution at such large $x$. This topic was partly discussed
in connection with a possible low-energy facility \cite{sk-rcnp}.
This example suggests the importance of the 50-GeV facility
by the following reason.

In order to find any new physics possibly beyond QCD,
we need to increase the ``resolution" $Q^2$ significantly.
Because the momentum fraction $x$ is given by $x=Q^2/2p\cdot q$,
for example in the lepton scattering, the large $Q^2$
roughly corresponds to large $x$. However, the large-$x$
parton distributions are not necessarily well known as obvious
from the interpretation of the above CDF events. 
What we have been doing first is to determine the parton distributions
at fixed $Q^2$ ($\equiv Q_0^2\sim$1 GeV$^2$) from various
high-energy-reaction data with typical
$Q^2=1\sim $a few hundred GeV$^2$. Then, DGLAP evolution equations
are used for calculating the variation of the distributions
from $Q_0^2$ to the large-$Q^2$ points, where the anomalous data are taken.
Therefore, it is crucial to determine accurate distributions
at $Q_0^2$ and especially at large $x$. Because present high-energy
accelerators focus inevitably on the small-$x$ region, they are not
advantageous to such physics. The 50-GeV facility should be
a unique one in studying the larger-$x$ region. 
We believe that the primary-beam experiments could have impact
on other fields of particle physics if it is properly used.

\vspace{0.1cm}
%%%%%%%%%%%%%%%%%%%%%%%%%%%%%%%%%%%%%%%%%%%%%%%%%%%%%%%%%%%%%%%%%%%%%%%%%%%%%%
%%%%%%%%%%%%%%%%%%%%%%%%%%%%%%%%%%%%%%%%%%%%%%%%%%%%%%%%%%%%%%%%%%%%%%%%%%%%%%
\section{Polarized proton-deuteron Drell-Yan process}
\label{pd-dy}
\vspace{-0.1cm}

Spin structure of the proton has been investigated mainly by polarized
lepton-nucleon scattering and will be also studied by polarized proton-proton
scattering at RHIC. There are already many data on the structure function
$g_1$ and we have rough idea on the polarized parton distributions
\cite{aac}. It is desirable to use different observables in order to
test our understanding of hadron spin structure. Additional spin structure
functions for the deuteron could be suitable quantities. Theoretically,
this topic has been investigated in the last ten years, and it is known
that there exists a new leading-twist structure function $b_1$ \cite{b1}.
Because it has not been measured at all, it should a good idea to
test theoretical predictions in comparison with future lepton scattering
data. In addition, a theoretical formalism had been completed recently
for the polarized pd Drell-Yan process \cite{hk}. The results suggested that
there exist many new structure functions which are associated with
the deuteron tensor structure. 
There are two major reasons for studying the polarized pd Drell-Yan 
process. The first purpose is, as mentioned above, to investigate
new structure functions which do not exist in the spin-1/2 proton.
The second one is to investigate antiquark flavor asymmetry
as discussed in Sec. \ref{flavor}. 

In this section, we explain the major consequences of the polarized 
pd formalism without discussing the details.
The polarized proton-proton (pp) Drell-Yan process has been investigated
theoretically for a long time and the studies are the basis of
the RHIC-Spin project. Reference \cite{hk} extended these studies to
the polarized pd Drell-Yan by taking into account the tensor structure
of the deuteron. 

A general formalism of the pd Drell-Yan was first studied in Ref. \cite{hk}
by using spin-density matrices and the Ralston-Soper type analysis.
Then, it was found that many new structure functions exist
due to the spin-1 nature of the deuteron.
The process was also analyzed in a quark model. The hadron tensor is
first written for an annihilation process
$q+\bar q \rightarrow \ell^+ + \ell^-$
by correlation functions. They are expanded in terms of
the sixteen $4\times 4$ matrices:
${\bf 1},\, \gamma_5,\, \gamma^\mu,\, \gamma^\mu \gamma_5,\, 
    \sigma^{\mu \nu} \gamma_5$ together with
kinematically possible vectors under the conditions of
Hermiticity, parity conservation, and time-reversal invariance.
We found in the analysis that there exists only one additional
spin asymmetry to the pp Drell-Yan case, and it was called
the unpolarized-quadrupole $Q_0$ asymmetry:
\begin{equation}
A_{UQ_0}  =  \frac{\sum_a e_a^2 \, 
                  \left[ \, f_1(x_1) \, \bar b_1(x_2)
                          + \bar f_1(x_1) \, b_1(x_2) \, \right] }
                {\sum_a e_a^2 \, 
                  \left[ \, f_1(x_1) \, \bar f_1(x_2)
                          + \bar f_1(x_1) \, f_1(x_2) \, \right] }
\, .
\label{eqn:auq0}
\end{equation}
Here, $f_1(x)$ and $\bar f_1(x)$ are unpolarized quark and
antiquark distributions, and $b_1(x)$ and $\bar b_1(x)$ 
are tensor-polarized distributions. The momentum fractions
are denoted as $x_1$ and $x_2$ for partons in the hadron
1 (proton) and 2 (deuteron), respectively.
This asymmetry is measured by using the unpolarized proton
and tensor polarized deuteron. It should provide us new
information on the tensor-polarized distributions because
the unpolarized distributions are well known in the proton
and deuteron. If the large-$x_F$ region is considered, 
Eq. (\ref{eqn:auq0}) becomes
\begin{equation}
A_{UQ_0} \textrm{(large $x_F$)} 
      \approx \frac{\sum_a e_a^2 \, f_1(x_1) \, \bar b_1(x_2)}
                   {\sum_a e_a^2 \, f_1(x_1) \, \bar f_1(x_2)}
\ \ \ \text{at large $x_F$} \, .
\end{equation}
This equation suggests that antiquark tensor distributions
should be obtained rather easily by the quadrupole spin asymmetry
in the polarized pd Drell-Yan.

The $x$ dependence of $b_1$ has been investigated in quark models.
The $b_1$ vanishes in any models with only the S-wave; therefore,
it should probe orbital-motion effects which are related to the
tensor structure. Because the tensor spin structure
is completely different from the present longitudinal spin physics,
experimental data should provide challenging information for
theorists. Furthermore, it has not been measured at all, so
that the proposed 50-GeV facility has an opportunity of significant
contributions to a new area of high-energy spin physics.

\vspace{0.1cm}
%%%%%%%%%%%%%%%%%%%%%%%%%%%%%%%%%%%%%%%%%%%%%%%%%%%%%%%%%%%%%%%%%%%%%%%%%%%%%%
%%%%%%%%%%%%%%%%%%%%%%%%%%%%%%%%%%%%%%%%%%%%%%%%%%%%%%%%%%%%%%%%%%%%%%%%%%%%%%
\section{Polarized light-antiquark flavor asymmetry}
\label{flavor}
\vspace{-0.1cm}

It became clear in the last ten years that the light-antiquark
distributions are not flavor symmetric \cite{udbar} by
the Gottfried-sum-rule violation and Drell-Yan experiments.
In particular, the Fermilab Drell-Yan experiments clarified
the $x$ dependence of $\bar u/\bar d$ by using the difference
between the pp and pd cross sections. In the same way, the
difference between the polarized pp and pd cross sections
should be useful for determining the flavor asymmetry in
polarized light-antiquark distributions. 

There could be two issues in extracting the longitudinal one
$\Delta \bar u/\Delta \bar d$ and the transversity one
$\Delta_T \bar u/\Delta_T \bar d$. First, there was no theoretical
formalism of the polarized pd Drell-Yan before Ref. \cite{hk},
so that we were not sure how to formulate the spin asymmetry
in terms of polarized distributions. Now, this issue has been
clarified \cite{hk} and we can discuss the cross section ratio
pd/pp by using the results. Second, nuclear corrections should
exist in the deuteron. However, they are not the essential part,
so that they are neglected in the present studies. If experimental
data are obtained in future, such corrections should be taken
into account carefully. 

If higher-twist effects are neglected in
the longitudinal and transverse spin asymmetries, the expressions
for the cross-section ratios are given by
\small
\begin{equation}
R_{pd} \equiv \frac{     \Delta_{(T)} \sigma_{pd}}
                   {2 \, \Delta_{(T)} \sigma_{pp}}
        =     \frac{ \sum_a e_a^2 \, 
    \left[ \, \Delta_{(T)} q_a(x_1) \, 
              \Delta_{(T)} \bar q_a^{\, d}(x_2)
            + \Delta_{(T)} \bar q_a(x_1) \, 
              \Delta_{(T)} q_a^d(x_2) \, \right] }
              { 2 \, \sum_a e_a^2 \, 
    \left[ \, \Delta_{(T)} q_a(x_1) \, 
              \Delta_{(T)} \bar q_a(x_2)
            + \Delta_{(T)} \bar q_a(x_1) \, 
              \Delta_{(T)} q_a(x_2) \, \right] }
\, ,
\label{eqn:ratio1}
\end{equation}
\normalsize
where $\Delta_{(T)}=\Delta$ or $\Delta_T$ depending on
the longitudinal or transverse case.
If the valence-quark distributions satisfy
$\Delta_{(T)} u_v (x \rightarrow 1) \gg 
 \Delta_{(T)} d_v (x \rightarrow 1)$ at large $x_F=x_1-x_2$, 
Eq. (\ref{eqn:ratio1}) becomes
\small
\begin{equation}
R_{pd} (x_F\rightarrow 1) = 1 - \left [ \,
    \frac{ \Delta_{(T)} \bar u (x_2) - \Delta_{(T)} \bar d (x_2) }
         { 2 \, \Delta_{(T)} \bar u (x_2) } \, \right ]_{x_2\rightarrow 0}
   =  \frac{1}{2} \, \left [ \, 1 
                 + \frac{\Delta_{(T)} \bar d (x_2)}
                        {\Delta_{(T)} \bar u (x_2)} 
                    \, \right ]_{x_2\rightarrow 0}
\, .
\label{eqn:rpd+1}
\end{equation}
\normalsize
Namely, the deviation from one indicates the difference
between $\Delta_{(T)} \bar u$ and $\Delta_{(T)} \bar d$ directly.
On the other hand, if another limit is taken $x_F\rightarrow -1$,
the ratio becomes
\small
\begin{equation}
R_{pd} (x_F\rightarrow -1) = 
     \frac{1}{2} \, \left [ \, 1 
                 + \frac{\Delta_{(T)} \bar d (x_1)}
                   {4 \, \Delta_{(T)} \bar u (x_1)} 
                    \, \right ]_{x_1\rightarrow 0}
\, .
\label{eqn:rpdm2}
\end{equation}
\normalsize
The factor of 1/4 suggests that the ratio of this region is
not as sensitive as the one of the large-$x_F$ region.

%%%%%%%%%%%%%%%%%%%%%%%%%%%%%%%% figure %%%%%%%%%%%%%%%%%%%%%%%%%%%%%%%%%%%%%%
\begin{wrapfigure}{r}{0.46\textwidth}
   \vspace{-0.4cm}
   \begin{center}
       \epsfig{file=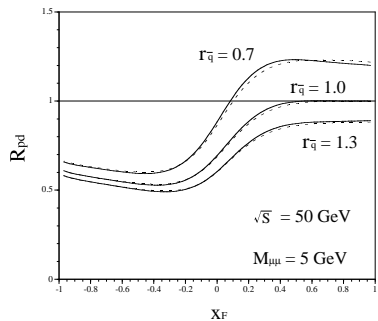,width=6.0cm}
   \end{center}
   \vspace{-0.7cm}
       \caption{\footnotesize
          Drell-Yan cross section ratio 
          $R_{pd} \equiv \Delta_{(T)}\sigma_{pd}/2 \Delta_{(T)}\sigma_{pp}$.
          (from Ref. [9]).}
       \label{fig:rpd}
\end{wrapfigure}
%%%%%%%%%%%%%%%%%%%%%%%%%%%%%%%% figure %%%%%%%%%%%%%%%%%%%%%%%%%%%%%%%%%%%%%%
We discuss numerical results for the ratio $R_{pd}$ in Fig. \ref{fig:rpd}
\cite{km}. For the parton distributions, we use a recent parametrization
in Ref. \cite{lss} at $Q^2$=1 GeV$^2$. The transversity distributions
are assumed to be the same as longitudinally-polarized ones.
Then, three ratios are assumed for the antiquark distributions:
$r_{\bar q} \equiv \Delta_{(T)} \bar u / \Delta_{(T)} \bar d$
=0.7, 1.0, or 1.3. Then, they are evolved to
$Q^2=M_{\mu\mu}^2$=25 GeV$^2$ by the LO-DGLAP evolution equations.
The numerical results indicate that the obtained ratios are much
different depending on the flavor-asymmetry ratio $r_{\bar q}$
particularly in the large-$x_F$ region. Therefore, the measurement
of this ratio could determine $r_{\bar q}$ and it should be an important
test of theoretical models for explaining the unpolarized flavor
asymmetry.

\vspace{0.1cm}
%%%%%%%%%%%%%%%%%%%%%%%%%%%%%%%%%%%%%%%%%%%%%%%%%%%%%%%%%%%%%%%%%%%%%%%%%%%%%%
%%%%%%%%%%%%%%%%%%%%%%%%%%%%%%%%%%%%%%%%%%%%%%%%%%%%%%%%%%%%%%%%%%%%%%%%%%%%%%
\section{Summary}\label{summary}
\vspace{-0.1cm}

We discussed first why the 50-GeV PS facility is important for
structure-function studies in the large-$x$ region.
Then, specific topics are discussed by using polarized deuteron.
We explained that the polarized proton-deuteron Drell-Yan process
is interesting in two respects.
First, it should be valuable for finding new tensor structure functions
in the deuteron. Second, it could be used for studying flavor asymmetry
in the polarized light-antiquark distributions. Because these topics are not
investigated by other facilities, possible 50-GeV data should
provide important information on hadron spin structure. 

\vspace{0.1cm}
%%%%%%%%%%%%%%%%%%%%%%%%%%%%%%%%%%%%%%%%%%%%%%%%%%%%%%%%%%%%%%%%%%%%%%%%%%%%%%
%%%%%%%%%%%%%%%%%%%%%%%%%%%%%%%%%%%%%%%%%%%%%%%%%%%%%%%%%%%%%%%%%%%%%%%%%%%%%%
\section*{Acknowledgments}
\vspace{-0.1cm}

S.K. was partly supported by the Grant-in-Aid for Scientific Research
from the Japanese Ministry of Education, Science, and Culture under
the contract number 10640277.
 
%%%%%%%%%%%%%%%%%%%%%%%%%%%%%%%%%%%%%%%%%%%%%%%%%%%%%%%%%%%%%%%%%%%%%%%%%%%%%%
\vspace{0.6cm}

\noindent
{* Email: kumanos@cc.saga-u.ac.jp. Information on his research
          is available at} \\

\vspace{-0.55cm}
\noindent
{\ \ \ http://www-hs.phys.saga-u.ac.jp.}  \\

\vspace{-0.4cm}
%\vspace{0.1cm}
%%%%%%%%%%%%%%%%%%%%%%%%%%%%%%%%%%%%%%%%%%%%%%%%%%%%%%%%%%%%%%%%%%%%%%%%%%%%%%%
%%%%%%%%%%%%%%%%%%%%%%%%%%%%%%%%%%%%%%%%%%%%%%%%%%%%%%%%%%%%%%%%%%%%%%%%%%%%%%%
%\section*{References}

\end{document}